\documentstyle[preprint,aps]{revtex}
\begin{document}
\tighten
\title{Nonradiative proton--deuteron fusion in stellar plasma}
\author{
S. A. Rakityansky\footnote{Permanent address: Joint
	Institute  for Nuclear Research,Dubna, 141980, Russia},
	S. A. Sofianos, L. L. Howell, M. Braun}
\address{Physics Department, University of South Africa,
	 P.O.Box 392, Pretoria 0001, South Africa}
\author{V. B. Belyaev}
\address{
Joint Institute  for Nuclear Research,Dubna, 141980, Russia}

\date{\today}
\maketitle
\begin{abstract}
The nuclear reaction $e+p+d\rightarrow {^3{\rm He}}+e$ is considered at
thermonuclear energies. The motion of the electron is treated within
the adiabatic approximation and the $pd$ scattering state is constructed
in the form of an antisymmetrized product of the bound state wave function
of the deuteron and of the  wave function of the $pd$ relative motion. The
latter is calculated using an effective $pd$ potential constructed via
the Marchenko inverse scattering method.
The bound state wave function of $^3{\rm He}$ is obtained using
Faddeev--type integrodifferential equations. The reaction rate thus
obtained for the solar interior conditions is approximately $10^{-4}$
of the corresponding rate for the radiative capture
      $pd\rightarrow{}^3{\rm He}\gamma$.
\end{abstract}
\newpage
\section{Introduction}
Burning of hydrogen in the main sequence stars mainly occurs through the
$pp$--chain which begins with the reaction $pp\longrightarrow e^+d\nu$.
It is generally accepted \cite{astro} that the second step of this chain
is the radiative capture of protons by deuterons,
\begin{equation}
\label{pdg}
	       p+d\longrightarrow {^3{\rm He}} +\gamma\,.
\end{equation}
However, due to the high densities in stars, the helium nuclei and other
intermediate products of the $pp$--chain can emerge not only from
two--body  but  from three--body initial states as well. Thus, for
example, $^3{\rm He}$ can be formed in the radiative capture
(\ref{pdg}) as well as in the three-body nonradiative fusion
process
\begin{equation}
\label{pde}
	     e +  p + d \longrightarrow {^3{\rm He}} + e\,.
\end{equation}
 The knowledge of the  production rates of various
nuclei in the stellar plasma is very important, not only in understanding
the production of energy  in stars but also in explaining  the abundance
of the elements observed in nature, and in
describing events during the first thousand seconds of the
evolution of the universe which were predominantly determined by the
nucleosynthesis of light  elements \cite{prim}.
The abundance of light elements in the Universe, together  with the
Hubble expansion and the relic backround radiation are experimental
evidences for the idea of the hot origin of the Universe in the Big
Bang process which in turn is closely related to the Grand Unification
and QCD theories \cite{prim,BBang}.\\

Values of the primordial abundance of light elements put some constraints
on the baryon density of the Universe as well as on the number of species
of light particles. For instance,  present theoretical estimates for the
primordial abundance of
       $d$, $^3{\rm He}$, $^4{\rm He}$, and $^7{\rm Li}$
can be in agreement with the corresponding experimental values only if the
number of the neutrino species is $N_\nu\le 3.9$ \cite{turner}.
Furthermore, from the theory of nucleosynthesis one can derive a
stringent limit to the existence of new light particles and even a
bound to the mass of the $\tau$--neutrino, namely, between 0.5 MeV
and 30 MeV \cite{kolb,dodelson}.\\

Any theory on the evolution of the universe  or nucleosynthesis must
deal with the total rates of nuclei production. In this respect the
thermonuclear reactions with two--body initial states, such as the reaction
(\ref{pdg}) have been  extensively  investigated \cite{cosmos}. However,
the role of the three--body mechanism in nucleosynthesis has not yet been
properly studied despite the fact that the three--body
processes have different selection rules and due to this it can have an
influence on the production of light nuclei.\\

The aim of this work is to estimate the relative significance of the
nonradiative process (\ref{pde}) in comparison with the radiative capture
(\ref{pdg}) in  the stellar $pp$--chain. In stellar plasma
nuclei are surrounded by electron gas  which has  a twofold
influence on  nuclear fusion processes. We can distinguish between
static and dynamic electronic effects \cite{leeb}. The former is the
screening of the Coulomb repulsion between the nuclei, and the latter
stems from the coupling between the electronic and nucleonic degrees of
freedom. Due to this coupling, energy and angular momentum
can be transferred from the nucleons to the electrons according to the
prevailing conservation laws. These dynamic electronic effects can
pave the way for a variety of possible fusion reactions in the plasma, which
otherwise are forbidden.\\

The nonradiative fusion (\ref{pde}) is an example where these effects are
manifested. The electron which is in the vicinity of the $pd$--pair while
they are interacting can carry away the excess energy, leaving the
three nucleons in a bound state. This is a kind of Auger transition in the
continuous spectrum. Since electrons move much faster than  nucleons, in
considering the scattering of an electron by a $pd$--pair,
its motion can be  treated within the adiabatic approximation.
In this approximation the nucleons are considered as being fixed at their
spatial positions during the electron scattering and thus the corresponding
amplitude will depend on the nucleon coordinates parametrically. The
physical amplitude can then be obtained by averaging over these coordinates
with the help of the wave functions describing the initial and final
configurations of the nucleons. To obtain the amplitude of the reaction
(\ref{pde}) we shall average the fixed--scatterer amplitude over the $pd$
scattering state and the $^3{\rm He}$ bound state.\\

The paper is organized as follows. In Sec. II we describe our formalism and
outline the procedure employed to evaluate the various  ingredients used to
obtain the reaction rate.  In  Sec. III we present our
results and conclusions. Some details concerning the derivation of
the reaction rate  formula are given in the Appendix.

\section{Formalism}

We are concerned with electron collisions with nuclei in a stellar plasma
consisting of protons, deuterons, and  electrons. Let $\Psi_{\bf k}$ be the
wave function of the relative motion of the $pd$--pair with momentum
${\bf k}$ and let ${\bf p}$  be the momentum of the electron with respect to
this pair.  The reaction rate for the collision process (\ref{pde}) per
unit volume per second is defined by \cite{gw}
\begin{equation}
\label{R}
      {\cal R}({\bf k},{\bf p}\longrightarrow{\bf p}')=\delta(E_f-E_i)
      (2\pi)^7\left|\langle\Psi_3,{\bf p}'|
      T|\Psi_{{\bf k}},{\bf p}\rangle\right|^2n_pn_dn_e
\end{equation}
where  the  states in the continuum are normalized as
$$
      \left\langle\Psi_{{\bf k}'},{\bf p}'|
      \Psi_{{\bf k}},{\bf p}\right\rangle=
      \delta({\bf k}'-{\bf k})\delta({\bf p}'-{\bf p})\,.
$$
$T$ is the transition operator, $\Psi_3$ is the bound state wave
function of $^3{\rm He}$ and  $n_p$, $n_d$, and $n_e$ represent particle
densities.\\

In stellar plasmas the momenta  ${\bf k}$ and ${\bf p}$ are distributed
according  to Maxwell's law
\begin{eqnarray}
\nonumber
    N_{\bf k}&=&(2\pi\mu\kappa\Theta)^{-3/2}
    \exp\left(-\displaystyle
    \frac{k^2}{2\mu\kappa \Theta}\right)\,,\\
\nonumber
    N_{\bf p}&=&(2\pi m\kappa \Theta)^{-3/2}
   \exp\left(-{\displaystyle \frac{p^2}{2m\kappa \Theta}}\right)\,,
\end{eqnarray}
where $N_{{\bf k}}$ and $N_{{\bf p}}$ are the probability densities, $\mu$
is the proton--deuteron reduced mass, $m$ is the electron mass, $\kappa$
is the Boltzmann constant, and $\Theta$ is the plasma temperature. Thus
 the reaction rate (\ref{R}) must be  averaged over the initial momenta
${\bf k}$ and ${\bf p}$ and integrated over the final momentum ${\bf p}'$,
i.e.,
\begin{equation}
\label{RA}
      \langle{\cal R}\rangle=\int\int\int d{\bf k} d{\bf p} d{\bf p}'
     \,{\cal R}({\bf k},{\bf p}\longrightarrow{\bf p}')N_{\bf k}N_{\bf p}\,.
\end{equation}

In what follows we shall discuss the various ingredients of this formula in
somewhat more detail.

\subsection{Transition operator}
Consider a four--body system consisting of an electron and three nucleons
described by the Jacobi vectors  shown in Fig. I.
The total  Hamiltonian of this system consists of three terms, namely,
\begin{equation}
\label{H}
	     H=H_3+h_0+V_e \, ,
\end{equation}
where $H_3$ is the total nuclear Hamiltonian, $h_0$ is the
kinetic energy operator for the free motion of the electron with respect to
the center of mass of the nucleons, and $V_e$ is the sum of the Coulomb
potentials between the electron and nucleons
\begin{equation}
\label{V}
     V_e=-\frac{\hat Z_1e^2}{r_1}-\frac{\hat Z_2e^2}{r_2}-
	 \frac{\hat Z_3e^2}{r_3} \,.
\end{equation}
Here $r_i$ are the distances between the electron and the nucleons, and
$\hat Z_i$ is the charge--operator for the $i$--th nucleon with
\begin{equation}
\label{Z}
	 \hat Z_i|1/2,\tau_i>=\cases {1, & if \quad $\tau_i=+1/2$ \cr
		  0,& if \quad $\tau_i=-1/2$ , \cr}
\end{equation}
where $|1/2,\tau_i>$ is the isotopic state of the $i$--th nucleon with
 $\tau_i$ being the third component of its isospin.\\

Since the motion of nucleons is much slower than that of the electron,
the $T$--matrix describing the electron scattering can be found with the
help of the fixed--scatterer approximation
\begin{equation}
\label{FSA}
	       T(z)=V_e+V_eG_0(z)T(z)\,,
\end{equation}
where $z$ is the total energy and
$$
	       G_0(z)=(z-h_0)^{-1}\,,
$$
is the free Greens' function.
Parametrical dependence of the potential (\ref{V}) on the nuclear Jacobi
vectors $\left\{\mbox{\boldmath $\rho$},{\bf r}\right\}$ and on the nuclear
isospin state $|\eta>$ of the three nucleons, makes the
fixed--scatterer $T$--matrix (\ref{FSA}) also  parametrically dependent
on them,
$$
      \langle\mbox{\boldmath $\rho'$},{\bf r}';\eta';{\bf p}'|T(z)|
   \mbox{\boldmath $\rho$},{\bf r};\eta; {\bf p}\rangle=
    \delta_{\eta\eta'}\delta(\mbox{\boldmath $\rho'$}-
   \mbox{\boldmath $\rho$})\delta({\bf r}'- {\bf r})
     T_{{\bf p}'{\bf p}}^\eta(\mbox{\boldmath $\rho$},{\bf r};z)\,,
$$
where ${\bf p}$ and ${\bf p}'$ are the initial and final momenta of the
electron. The physical $T$--matrix is obtained by the averaging,
\begin{equation}
\label{AVERAGE}
       \langle\Psi_3,{\bf p}'|T(z)|\Psi_{\bf k},{\bf p}\rangle=
       \sum_{\chi\eta}
	\int\int d\mbox{\boldmath $\rho$} d{\bf r}\,
	{\psi_3^{\chi\eta}}^*(\mbox{\boldmath $\rho$},{\bf r})
	T_{{\bf p}'{\bf p}}^\eta(\mbox{\boldmath $\rho$},{\bf r};z)
	\psi_{{\bf k}}^{\chi\eta}(\mbox{\boldmath $\rho$},{\bf r})
	\,,
\end{equation}
where  $\psi_{{\bf k}}^{\chi\eta}$ and $\psi_3^{\chi\eta}$  are the spatial
parts of the three--nucleon wave functions corresponding to the
spin--isospin state $ |\chi\eta>$,
\begin{equation}
	|\chi\eta>\equiv
      |((s_1s_2)s_{12},s_3)\displaystyle\frac{1}{2}\sigma;
	((t_1t_2)t_{12},t_3)\displaystyle\frac{1}{2}\tau\rangle
\label{xieta}
\end{equation}
i.e.,
\begin{equation}
     \psi_{{\bf k}}^{\chi\eta}(\mbox{\boldmath $\rho$},{\bf r})
     =\langle \mbox{\boldmath $\rho$},{\bf r};\chi\eta|\Psi_{{\bf k}}
      \rangle\,,
\label{cont}
\end{equation}
and
\begin{equation}
     \psi_3^{\chi\eta}(\mbox{\boldmath $\rho$},{\bf r})
     =\langle \mbox{\boldmath $\rho$},{\bf r};\chi\eta|\Psi_3
      \rangle\,.
\label{bound}
\end{equation}

\subsection{Proton--deuteron scattering state}
The thermonuclear energies we are considering are far below the threshold
of the deuteron break--up and we can  therefore neglect the three--nucleon
continuous spectrum $(1+1+1)$ and construct the $(1+2)$ scattering state in
the form of the antisymmetrized product of the proton--deuteron relative
motion wave function $\varphi_{\bf k}$ and the deuteron wave function
$\psi_2$,
$$
   \Psi_{\bf k}={\cal A}\left\{\varphi_{\bf k}\psi_2\chi\eta\right\}\,.
$$
Since the deuteron wave function is antisymmetric,
the antisymmetrizer $\cal A$ involves only the permutations
$P_{23}$ and $P_{13}$,
\begin{equation}
\label{A}
    {\cal A}={\displaystyle\frac{1}{\sqrt 3}}\left(1-P_{23}-P_{13}\right)\,.
\end{equation}
Both $\varphi_{\bf k}$ and  $\psi_2$ are assumed to
have $S$--wave components only with the total spin of the  $nd$--system
equal to $1/2$. Before the antisymmetrization, the spin--isospin
states of the $nd$--system are
      $|\chi\eta\rangle=|((s_1s_2)1s_3)\frac12\sigma;
       ((t_1t_2)0t_3)\frac12\tau \rangle$.
Thus after the nucleon permutations we obtain
\begin{eqnarray}
\label{PSIKA}
   \langle\mbox{\boldmath $\rho$},{\bf r}|\Psi_{\bf k}\rangle&=&
       {\displaystyle\frac{1}{\sqrt 3}}\left\{
   \varphi_{\bf k}(\mbox{\boldmath $\rho$})
       \psi_2({\bf r})
       \left|\left.((s_1s_2)1s_3)
       \frac12\sigma; ((t_1t_2)0t_3)\frac{1}{2}\tau
	\right\rangle\right.\right.\\
\nonumber
    &-&\varphi_{\bf k}(-\frac12\mbox{\boldmath $\rho$}+\frac34{\bf r})
     \psi_2(\mbox{\boldmath $\rho$}+
     \frac{1}{2}{\bf r})\left|\left.((s_1s_3)1s_2)\frac12\sigma;
     ((t_1t_3)0t_2)\frac12\tau\right\rangle\right.\\
\nonumber
    &-&\left.\varphi_{\bf k}(-\frac12\mbox{\boldmath $\rho$}-\frac34
    {\bf r})\psi_2(\frac12{\bf r}-
     \mbox{\boldmath $\rho$})\left|\left.((s_3s_2)1s_1)\frac12\sigma;
    ((t_3t_2)0t_1)\frac12\tau \right \rangle \right. \right\}\,.
\end{eqnarray}
The spatial components
   $  \psi_{{\bf k}}^{\chi\eta}(\mbox{\boldmath $\rho$},{\bf r})$,
   Eq. (\ref{cont}), can be obtained by projecting on to the spin-isospin
   states  (\ref{xieta}).\\

To find the deuteron wave function $\psi_2({\bf r})$ we solve the  two--body
Schr\"odinger equation with the  Malfliet--Tjon I-III \cite{mt}
$NN$--potential. In order to obtain the relative motion wave function
$\varphi_{{\bf k}}(\mbox{\boldmath $\rho$})$ the two--body scattering problem
is solved using the Jost function  method proposed in Ref. \cite{RSA96} with
an effective proton--deuteron potential $V_{pd}(\rho)$ which consists of
two terms:
\begin{equation}
  V_{pd}(\rho)=V_c(\rho)\exp\left(-\frac{\rho}{\rho_D}\right)+V_s(\rho).
\label{Vpd}
\end{equation}
The first term describes the  proton--deuteron
Coulomb interaction. Since deuteron is not a point--like particle, we take
into account the spherically symmetric distribution of its charge in space
by using
$$
      V_c(\rho)=\frac{4\pi e^2}{\rho}\int_0^{2\rho}dr\,\left[
      r\psi_2(r)\right]^2\,.
$$
 At large $\rho$ (beyond the deuteron radius),
$V_c(\rho)$ coincides with the simple Coulomb potential $e^2/\rho$. However
at small $\rho$  it behaves quite differently. In particular, $V_c(\rho)$ is
not singular  at $\rho=0$, but instead $V_c(0)=0$. The exponential
factor  stems from electron screening and follows from the standard
Debye--H\"uckel theory \cite{salpeter}. For the Debye radius we use the
value
$$
    \rho_D=21800\,\,{\rm fm}
$$
which corresponds to the solar plasma conditions and is typical for other
stars \cite{leeb}.\\

The $V_s(\rho)$ is the strong (nuclear) $pd$ interaction and we constructed
it using the $\ell$--~dependent Marchenko inverse scattering method
\cite{AM63,CS77} which we briefly describe next.\\

\subsection{Marchenko Inverse Scattering method}\label{Mar}
In the Marchenko inverse scattering method a unique, energy--independent,
$\ell$--dependent, local potential $V_\ell(\rho)$ can be constructed
which is phase equivalent to the $nd$ doublet channel
effective interaction at all energies. This potential is obtained from
\begin{equation}
      V_\ell(\rho) = - 2\frac{\rm d}{d\rho} K_\ell(\rho,\rho)
\label{velp}
\end{equation}
where the kernel $K_\ell(\rho,\rho')$ obeys the Marchenko  fundamental
equation
\begin{equation}
     K_\ell(\rho,\rho') + F_\ell(\rho,\rho') +\int_r^\infty
     K_\ell(\rho,\rho'') F_\ell(\rho'',\rho')d\rho'' = 0.
\label{march}
\end{equation}
Here, $\rho$ is the relative distance between the neutron and deuteron
and is canonically conjugate to the relative momentum $k$. The driving
term $F_\ell(\rho,\rho')$ is given by
\begin{equation}
       F_\ell(\rho,\rho') = \frac{1}{2\pi}
	  \int_{-\infty}^{+\infty} w_\ell^+(k\rho)
	  [1-S_\ell(k)] w_\ell^+(k\rho'){\rm d}k
	  + A_\ell w_\ell^+(b_{\ell}\rho) w_\ell^+(b_{\ell}\rho').
\label{fl}
\end{equation}
$S_\ell(k)$ is the S--matrix for the specific partial wave
$\ell$, and the function $w_\ell^+(z)$ is related to the spherical
Hankel function $h_\ell^{(+)}(z)$ by
\begin{equation}
       w_\ell^+(z) = i e^{i\pi\ell}z h_\ell^{(+)}(z). \label{wl}
\end{equation}
Furthermore, $A_\ell$ is the so--called asymptotic bound state
normalisation constant, while
$b_{\ell} = \sqrt{ - 4 M E_{b}^{\ell}/3}$, $E_{b}^{\ell}$
being the three--body bound state energy and $M$ the nucleon mass.\\

As is apparent from its definition (\ref{fl}) the evaluation
of $F_\ell(\rho,\rho')$ requires the knowledge of the S--matrix for all
real energies from the elastic scattering threshold to infinity, together
with the reflection property $S_\ell(- k)= 1/S_\ell(k)$, as well as the
binding energy and the corresponding asymptotic bound state normalisation
constant. It is greatly simplified by choosing a rational (Bargmann--type)
parametrisation
\begin{equation}
     S_\ell(k) = \frac{k+ib_\ell}{k-ib_\ell}\prod_{n=1}^{N_{\ell}}
     \frac{k+\alpha_{n}^{\ell}}{k-\alpha_{n}^{\ell}} \label{sl}
\end{equation}
The number $N_{\ell}$ is to be taken odd to satisfy the
requirement $S_\ell\rightarrow 1$ for $k\rightarrow 0$.
The $\alpha_{n}^{\ell}$ are complex numbers used to fit
the (numerically) given S--matrix.\\

With the choice (\ref{sl}) the integration in Eq. (\ref{fl})
can be easily performed analytically, the result being
\begin{equation}
    F_\ell(\rho,\rho') = -i\sum_{m=1}^{N_{\ell}^u}
      R_m^{\ell} w_\ell^+
     (\alpha_m^{\ell}\rho)w_\ell^+(\alpha_m^{\ell}\rho')
     + A_\ell w_\ell^+(b_{\ell}\rho) w_\ell^+(b_{\ell}\rho')\,.
\label{fl2}
\end{equation}
Here, $R_m^{\ell}$ are the coefficients of the residues of the
integrand for the $N_{\ell}^u$ poles of (\ref{sl}) which are lying in
the upper  half complex k--plane (excluding the one corresponding to
the bound state at $k = ib_\ell$). The separable form of
$F_\ell (\rho,\rho')$ provides us with an algebraic solution
of the integral equation (\ref{march}) for the kernel $K_\ell (\rho,\rho')$
from which the potential can be obtained via Eq. (\ref{velp}).\\

In order that the potential be unique, the  asymptotic bound state
normalisation constant $A_\ell$ must be given a definite value. We
choose it to be  \cite{CS77}
\begin{equation}
    A_\ell = i \frac{ f_\ell (-ib_\ell)}{ f^\prime_\ell (ib_\ell)},
    \label{al}
\end{equation}
where $f_\ell$ is the Jost function, and $ f^\prime_{\ell}(k) =
df_\ell (k)/ dk$.

\subsection{Wave function of\mbox{ $^3{\rm He}$}}\label{wf}
The integrodifferential equation approach to few- and many body systems
developed  by Fabre de la Ripelle and collaborators
\cite{Fabr1,IDEA3,IDEA4} is used to construct the bound state wave
function of  $^3{\rm He}$. In this method the potential $U$ is written
as  a sum of two--body  interactions
\begin{equation}
     U=\sum_{i<j} V(r_{ij})\,,
\end{equation}
and the three-body bound state wave function  is written as a sum
of two--body amplitudes
\begin{equation}
    \psi_3=H_{[L_m]}({\bf x})\sum_{i<j} F(r_{ij},r_0)\,,
\end{equation}
where $H_{[L_m]}({\bf x})$ is  the  harmonic polynomial of lowest
degree $[L_m]$ occuring in the harmonic polynomial expansion of the
wave function, and ${\bf x}$ represents the nucleon coordinates ${\bf x}_i$
with $ r_{ij}=|{\bf x}_i-{\bf x}_j|$,
and $r_0=[2/3\sum_{i<j} {\bf r}_{ij}^2]^{1/2}$ is the hyperradius.\\

With the above expansions one has to solve, instead of the Schr\"odinger
equation, the Faddeev--type equation for the amplitude $F(r_{ij},r_0)$
\begin{equation}
     (T-E)H_{[L_m]}({\bf x}) F(r_{ij},r_0)=- V(r_{ij})
    H_{[L_m]}({\bf x})\sum_{k<\ell} F(r_{k\ell},r_0)\,.
\label{TEH}
\end{equation}
A solution of this equation will be an approximate solution of the
Schr\"odinger equation for two--body amplitudes where pairs are in
$S$--states. Another Faddeev--type equation can be obtained by extracting
the hypercentral potential $V_{[L_m]}(r_0)$  of $V(r_{ij})$ \cite{Fabr1}
and by writing
\begin{eqnarray}
\label{TEH1}
     (T+\frac{A(A-1)}{2}V_{[L_m]}(r_0)&-&E)H_{[L_m]}({\bf x}) F(r_{ij},r_0)\\
      &=&-[V(r_{ij})-V_{[L_m]}(r_0)]
    H_{[L_m]}({\bf x})\sum_{k<\ell} F(r_{k\ell},r_0)\,.
\nonumber
\end{eqnarray}
This equation forms the basis of the integrodifferential equation approach
 (IDEA) to few-- and many--body systems \cite{IDEA1} and takes into account
in an
approximate way, via the hypercentral potential $V_{[L_m]}(r_0)$, the
effects of the coupling between  the orbitals $\ell\ne 0$ of the spectator
particle and the interacting pair. For $[L_m]=0$ then
$H_{[L_m]}({\bf x})=1$.
We notice that  by summing over all pairs one generates the  Schr\"odinger
equation but here the two--body potential is the residual interaction
on the right hand side of (\ref{TEH1}).\\

We assume that we have a central
spin--dependent nucleon-nucleon potential of the form
$$
   V^+(r_{ij},\sigma,\tau) =V^{1+}(r_{ij})P^{1+}_{ij}
	    + V^{3+}(r_{ij})P^{3+}_{ij}
$$
where the projection operators $P^{1+}_{ij}$ and $P^{3+}_{ij}$ are acting on the
singlet-- and triplet--even states respectively. In order to proceed
two further steps are required. In the first step  $F(r_{ij},r_0)$ is
written,
$$
     F(r_{ij},r_0 )=P(\zeta_{ij},r_0)/r_0^{(D-1)/2},
$$
where $\zeta_{ij}=2r^2_{ij}/r_0^2-1$ and $D=3A-3$.
In the second step Eq. (\ref{TEH1}) is projected on the
$r_{ij} $ space to give two coupled integrodifferential equations
\begin{eqnarray}
\nonumber
      \biggl [ \frac{\hbar^2}{m} \nabla^2_0 -\frac{A(A-1)}{2}V_0(r_0)+E
      \biggr ]\,P_0^S(\zeta,r_0)
     &=&
\biggl[ \frac{V^{1+}+V^{3+}}{2}-V_0(r_0)\biggr ]
      \Pi_0^{S}(\zeta,r_0)\\
\nonumber
      &+&\left[\frac{V^{1+}-V^{3+}}{2}\right]
      \Pi_0^{S'}(\zeta,r_0)\\
\label{IDEAS}
&&\\
\nonumber
      \biggl [ \frac{\hbar^2}{m} \nabla^2_0 -\frac{A(A-1)}{2}V_0(r_0)+E
      \biggr ]\,P_0^{S'}(\zeta,r_0)
     &=&
\biggl[ \frac{V^{1+}+V^{3+}}{2}-V_0(r_0)\biggr ]
      \Pi_0^{S'}(\zeta,r_0)\\
\nonumber
      &+&\left[\frac{V^{1+}-V^{3+}}{2}\right]
      \Pi_0^{S}(\zeta,r_0)\,,
\end{eqnarray}
where
\begin{equation}
\Pi_0^n(\zeta,r_0)=P_0^{n}(\zeta,r_0)+
	    \int_{-1}^{+1}f_{(0)}^{n}(\zeta,
  \zeta^\prime)P_0^{n}(\zeta^\prime,r_0)d\zeta^\prime
\end{equation}
with $n=S,S'$.

The $\nabla^2_0$ is given by
\begin{equation}
	  \nabla^2_0=\frac{\partial^2}{\partial r_0^2}
	  -\frac{{\cal L}_0({\cal L}_0+1)}
	  {r_0^2}+\frac{4}{r_0^2}\frac{1}{W_0(\zeta)}
	  \frac{\partial}{\partial \zeta}(1-\zeta^2)
	  \frac{\partial}{\partial \zeta},
\end{equation}
with ${\cal L}_0=(D-3)/2$, while the weight function $W_0$
by
\begin{equation}
	W_0(\zeta)=(1-\zeta)^{(D-5)/2}(1+\zeta)^{1/2}.
\end{equation}
The $V^{(i)}\equiv V^{(i)}\left(r_0\sqrt{(1+\zeta)/2}\right)$,
$i=1+$, $3+$,
are the even singlet and triplet nucleon--nucleon potentials.
The kernels $f_{(0)}^{n}(\zeta,\zeta^\prime)$, $n= S,S'$ result from
the   projection on to the $r_{ij}$ space.
More details concerning these kernels  and other technical
points can be found in Refs. \cite{IDEA1,IDEA3}.\\

Once the components $P_{ij}^{n}(\zeta,r_0)$ are found, one can construct
the various symmetries for the bound state wave function $\Psi_3$
 (\cite{IDEA4}). However, in the present
calculation we used only the space symmetric part i.e.,
$$
\langle\mbox{\boldmath $\rho$},{\bf r}|  \Psi_3\rangle =
     {\displaystyle \frac{\psi_3(\mbox{\boldmath $\rho$},
      {\bf r})}{\sqrt{2}}}\left[
     |((s_1s_2)1s_3)\frac12\sigma;((t_1t_2)0t_3)\frac12\tau\rangle-
     |((s_1s_2)0s_3)\frac12\sigma;((t_1t_2)1t_3)\frac12\tau\rangle\right]\,.
$$
It is emphasised here that the IDEA for the three--body case and
for S--projected potentials, is equivalent to the exact Faddeev equations.\\


\subsection{Electron scattering}
The scattering of the electron on the $i$--th nucleon, is described by
the $t$--matrix
\begin{equation}
	  t_i(z)=V_i+V_iG_0(z)t_i(z)\,,
\label{ti}
\end{equation}
where
$$
	   V_i=-\hat Z_ie^2/r_i\,,
$$
and $G_0(z)$ is the free Green function. Using (\ref{ti}) we can  rewrite
 Eq. (\ref{FSA}) as a sum of Faddeev components, viz.,
\begin{equation}
   T=\sum_{i=1}^3 T_i
\end{equation}
with
\begin{equation}
   T_i=t_i+t_iG_0\sum_{j\ne i} T_j
\end{equation}
Iterative solution of these equations provides  the multiple scattering
series
\begin{equation}
\label{SER}
     T=t_1+t_2+t_3+t_1G_0t_2+t_2G_0t_1+t_1G_0t_3+t_3G_0t_1+
     t_2 G_0t_3+t_3G_0t_2+\cdots\,.
\end{equation}
Since the average energy of the Coulomb interaction of the electron
with the nucleons is of atomic order of magnitude
$(\sim 10\,{\rm eV})$ and the average collision energy is
 $\langle\kappa\Theta\rangle \sim 10^3\,{\rm eV}$,
we can omit the higher order rescattering terms in Eq. (\ref{SER}),
that is, we use the following approximation \cite{Taylor}
\begin{equation}
\label{TAPPR}
       T\approx t_1+t_2+t_3\,.
\end{equation}
For the same reason we can apply the Born approximation for the
Coulomb $t$--matrix, i.e.
$$
    t_c({\bf p}',{\bf p};Q)\approx
       -\frac{Qe^2}{2\pi^2\left({\bf p}-{\bf p}'\right)^2}\,,
$$
where $t_c$ is the two--body Coulomb $t$--matrix for an electron scattered
off the charge $Qe$.\\

To obtain matrix elements of the $T$--operator (\ref{TAPPR}), we use the
following basis states
\begin{equation}
\label{BASIS}
      \left.\left|\mbox{\boldmath $\rho$},{\bf r},{\bf p};\chi
	\eta\right.\right\rangle\equiv
       \left.\left|\mbox{\boldmath $\rho$},{\bf r},{\bf p};
      ((s_1s_2)s_{12}s_3)\frac12\sigma;((t_1t_2)t_{12}t_3)\frac12\frac12
      \right.\right\rangle\,,
\end{equation}
where ${\bf p}$ is the electron momentum with respect to the center of mass
of the nucleons. Using the space displacement operators \cite{gw}
$\exp(-i\mbox{\boldmath $\Delta$}_i {\bf p})$,
where $\mbox{\boldmath $\Delta$}_i $ is a vector directed from
the center of mass to the $i$--th nucleon, we also construct the
shifted basis states,
$$
    \left.\left|\mbox{\boldmath $\rho$},{\bf r},{\bf p}_i;
      \chi\eta\right.\right\rangle=
     \exp(-i\mbox{\boldmath $\Delta$}_i{\bf p})
     \left.\left|\mbox{\boldmath $\rho$},{\bf  r},
     {\bf  p};\chi\eta\right.\right\rangle\,,
$$
which differ from (\ref{BASIS}) in the sense that the
electron has the same momentum ${\bf p}_i={\bf p}$ but with respect to the
$i$--th nucleon (not to the center of mass). Thus, we obtain
\begin{equation}
      \left\langle\mbox{\boldmath $\rho'$},{\bf r}',
       {\bf p}';\chi'\eta'\left|T(z)
       \right|\mbox{\boldmath $\rho$},
      {\bf r},{\bf p};\chi\eta\right\rangle =
       \delta_{\chi'\chi}\delta{\eta'\eta}
       \delta(\mbox{\boldmath $\rho'$}-\mbox{\boldmath $\rho$})
       \delta({\bf r}'-{\bf r})
       T_{{\bf p}'{\bf p}}^\eta(\mbox{\boldmath $\rho$},{\bf r};z)
\label{FSAT}
\end{equation}
with
\begin{equation}
       T_{{\bf p}'{\bf p}}^\eta(\mbox{\boldmath $\rho$},{\bf r};z)=
	\sum_{m=1}^3
	\exp\left[i({\bf p}-{\bf p}')\mbox{\boldmath $\Delta$}_m\right]
	t_c({\bf p}',{\bf p}; Q_m^\eta)\,,
\end{equation}
where the effective charges are
$$
  Q_m^\eta=\langle\eta\left|\hat Z_m\right|\eta\rangle\,.
$$
Using the $6j$--symbols \cite{war} for recoupling the isospins, we obtain
\begin{eqnarray}
\nonumber
       Q_1^0&=&\frac12\,,\qquad Q_2^0=\frac12\,,\qquad Q_3^0=1\,,\\
\nonumber
       Q_1^1&=&\frac56\,,\qquad Q_2^1=\frac56\,,\qquad Q_3^1=\frac13\,,
\end{eqnarray}
where the upper index $0$ designates the
       $|((t_1t_2)0t_3)\frac12\frac12\rangle$
and $1$ the
      $|((t_1t_2)1t_3)\frac12\frac12\rangle$
states.\\


\section{Results and conclusions}
The effective $pd$ potential  obtained  using the Marchenko inverse
scattering theory, described in Sec.  \ref{Mar}, for the $\ell=0$
partial wave,  is plotted in Fig. II. This potential  reproduces the
experimental $nd$ phase shifts as the Coulomb $pd$ interaction
has been treated separately as described in Eq. (\ref{Vpd}). Since these
phase shifts, however,  are available only at low energies we use for large
values of $k$  the phase shifts obtained via the
Faddeev equations \cite{Alt}.
The oscillations  in the interaction region are due to the opening of
the break--up channel as well as to the behaviour
of the phase shifts at large $k$ values ($k\ge 1000$ MeV). \\

The wave function of the $^3{\rm He}$--nucleus was obtained by solving
the system (\ref{IDEAS}) with the  Malfliet-Tjon I-III (MT I-III)
nucleon-nucleon potential \cite{mt} as input. The binding energy
obtained is 8.86 MeV  while the  root mean square radius is 1.685 fm.
However, for our final calculations we used the energy release
in the reaction (\ref{pde}), viz.,
$$
       {\cal E}=E_3-E_2\,,
$$
correponding to the difference between the experimental binding energies
of the deuteron and $^3{\rm He}$,  namely $E_2=$2.224574 MeV
and $E_3=$7.718109 MeV \cite{Waps}.\\

The rate of the reaction (\ref{pde}) can be presented in the following
factorized form (see the Appendix)
$$
    \langle{\cal R}\rangle=n_pn_dn_e\langle \Sigma \rangle\,,
$$
where the quantity $\langle \Sigma\rangle $  is
analogous to $\langle\sigma  v\rangle $ generally
used in  two--body reaction theories  \cite{cosmos}.
The calculated rates for different temperatures of the plasma are
presented in Table I, where the results are given
in units of ${\rm cm}^6{\rm mole}^{-2}{\rm sec}^{-1}$  \cite{fowler}.
These units are obtained when instead of
$n_pn_dn_e$ we multiply $\langle\Sigma\rangle$ by $N_A^2$, i.e. the Avogadro
number squared. The second column of the Table I contains the
values of
$$
{\cal R}_e=N_A^2\langle\Sigma\rangle
$$
while the third contains the  rate for
the radiative capture  (\ref{pdg}) which is also presented in the form
$$
{\cal R}_\gamma=N_A\langle\sigma  v\rangle \,,
$$
instead of $n_pn_d\langle\sigma v\rangle $,
that is in  ${\rm cm}^3{\rm mole}^{-1}{\rm sec}^{-1}$ units
(the data are taken from Ref. \cite{fowler}).\\

In order to assess the importance of the nonradiative process (\ref{pde})
as compared to the radiative capture (\ref{pdg})  when $^3{\rm He}$ nuclei
are  generated, we must compare the rates for these two processes
per ${\rm cm}^3$ per ${\rm sec}$. This requires the knowlege
of  the particle densities $n_p$,
$n_d$, and $n_e$, and therefore we need to specify the plasma conditions.
However, by considering the ratio of the two rates, viz.,
$$
   {\rm ratio}= n_e\frac{\langle\Sigma\rangle}{\langle\sigma v\rangle}\,.
$$
only one unknown  parameter remains, namely the electron density $n_e$.
We calculated this ratio for the reaction rates for (\ref{pde}) and
(\ref{pdg}) by using the value $n_e=100N_A\,{\rm cm}^{-3}$  corresponding
to the solar interior plasma \cite{astro}.
The calculated ratio is given in the fourth column of  Table I.
The ratio for any other electron density $n_e$ can be obtained by simply
multiplying these  values by $n_e/100N_A$.\\

It is seen that in the solar $pp$--chain the nonradiative fusion (\ref{pde})
 plays, apparently, a minor role. However, at the early stages of the
universe when $n_e/N_A \gg 100$, this reaction must have been
significant.

\bigskip
\noindent
{\Large \bf Acknowledgements}\\
One  of us (S.A.R)  gratefully acknowledges  financial support
from the University of South Africa and
the Joint Institute  for Nuclear Research, Dubna.

\newpage
\appendix
\section{Reaction Rate: Explicit formula}
Due to the very low nuclear energies involved we shall use only $S$--wave
components of the wave functions describing the deuteron, $\psi_2$, the
$pd$ relative motion, $\varphi_{\bf k}$, and the nucleus $^3{\rm He}$,
$\psi_3$. These functions depend only on the moduli of the
corresponding vector variables. However, due to the antisymmetrization,
 the dependence  on the angle between the vectors
$\mbox{\boldmath $\rho$}$ and ${\bf r}$ is also present.\\

The fixed scatterer $T$--matrix (\ref{FSA}) depends on all four
angles that define the directions of the vectors $\mbox{\boldmath $\rho$}$
and ${\bf r}$ which are involved in
(\ref{FSAT}) in the form of the following combinations
\begin{eqnarray}
\nonumber
     \mbox{\boldmath $\Delta$}_1&=&\frac12{\bf r}+\frac13
	   \mbox{\boldmath $\rho$}\,,\\
\nonumber
    \mbox{\boldmath $\Delta$}_2&=&\frac13\mbox{\boldmath $\rho$}
		  -\frac12{\bf r}\,,\\
\nonumber
      \mbox{\boldmath $\Delta$}_3&=&-\frac23\mbox{\boldmath $\rho$}\,.
\end{eqnarray}
Thus, the transition matrix (\ref{AVERAGE}) is a a
6--dimensional integral. The final averaging procedure involves 9 more
integrations,
$$
  \langle{\cal R}\rangle=n_pn_dn_e(2\pi)^7\int\int\int
    d{\bf k} d{\bf p} d{\bf p}'\,
   \delta(E_f-E_i)\left|\left\langle\Psi_3,{\bf p}'|T|\Psi_{\bf k},{\bf p}
   \right\rangle\right|^2N_{\bf k}N_{\bf p}\,.
$$
The transition matrix
   $\left\langle\Psi_3,{\bf p}'|T|\Psi_{\bf k},{\bf p}\right\rangle$
depends on $({\bf p}'-{\bf p})$ and $|{\bf k}|$. Moreover the Maxwell
distributions are isotropic. Hence, the 5--dimensional integration over
$d\Omega_{{\bf k}} d\Omega_{{\bf p}'} d\phi_{{\bf p}}$ is trivial and
results in the factor $32\pi^3$. One more integration, over the final
momentum, is performed with the help of the energy conserving
$\delta$--function and the 9--dimensional integral is reduced
to a 3--dimensional one. Thus, finally we obtain
$$
     \langle{\cal R}\rangle=n_p n_d n_e \langle\Sigma\rangle\ ,
$$
where $\langle\Sigma\rangle$ is defined by
\begin{eqnarray}
\nonumber
   \langle\Sigma\rangle&=&\frac{4\pi^3e^4c}
   {3\sqrt{m\mu^3}\kappa^3\Theta^3}
    \int_0^\infty \int_0^\infty  dkdp\,k^2p^2\exp\left(
   -\frac{k^2}{2\mu\kappa \Theta}-\frac{p^2}{2m\kappa \Theta}\right)\\
\nonumber
&\times&\int_{-1}^{+1}dx\,\frac{Q}{q^4}
      \left|
     \int_0^\infty  \int_0^\infty dr d\rho\,\int_0^{2\pi}\int_0^{2\pi}
     d\phi_r
	d\phi_\rho \int_{-1}^{+1}\int_{-1}^{+1}dy dz\,r^2
      \rho^2\psi_3(\rho,r)\right.\\
\nonumber
 & \times&
     \left\{\left[\varphi_k\left(\sqrt{
 9r^2/16+\rho^2/4-3{\bf r}\mbox{\boldmath $\cdot\rho$}/4}\ \right)
 \psi_2\left(\sqrt{r^2/4+\rho^2+{\bf r}\mbox{\boldmath $\cdot\rho$}}
     \ \right) \right. \right.\\
\nonumber
&+&\left.
      \varphi_k\left(\sqrt{9r^2/16+\rho^2/4
     +3{\bf r}\mbox{\boldmath $\cdot\rho$}/4}\ \right)
    \psi_2\left(\sqrt{r^2/4+\rho^2-
     {\bf r}\mbox{\boldmath $\cdot\rho$}}\ \right)\right]\\
\nonumber
&\times&
\left[3\exp\left(\frac{iq\rho y}{3}+\frac{iqr z}{2}\right)+
      3\exp\left(\frac{iq\rho y}{3}-\frac{iq r z}{2}\right)+
      2\exp\left(\frac{-i2q\rho y}{3}\right)\right]\\
\nonumber
&+&
\left.\left.\left.\left.\left.
   \varphi_k(\rho)\psi_2(\rho)
    \left[
      2\exp\left(\frac{iq\rho y}{3}+\frac{iq r z}{2} \right)
     + 2\exp \left(\frac{iq\rho y}{3}-\frac{iq r z}{2}\right)
      +4\exp\left(\frac{-i2q\rho y}{3}\right)
     \right]
\right.\right.\right.\right\}\right|^2\,,
\label{fc}
\end{eqnarray}
where the scalar product ${\bf r}\mbox{ \boldmath$\cdot\rho$}$ is
$$
   {\bf r}\mbox{ \boldmath$\cdot\rho$}=r\rho\left
     [\sqrt{(1-y^2)(1-z^2)}\cos(\phi_r-\phi_\rho)
+yz\right]\,,
$$
while
$$
      Q=\sqrt{2m{\cal E}+\frac{m}{\mu}k^2+p^2}\,,
$$
and
$$
      q=\sqrt{Q^2+p^2-2Qpx}\,.
$$


\newpage

\newpage


\def\emline#1#2#3#4#5#6{%
       \put(#1,#2){\special{em:moveto}}
       \put(#4,#5){\special{em:lineto}}}
\def\newpic#1{}
\begin{figure}
\caption{The Jacobi coordinates for the $e+p+d$ system.}
\vspace{2cm}
\unitlength=1.00mm
\special{em:linewidth 0.4pt}
\linethickness{0.4pt}
\begin{picture}(121.33,116.33)
\put(30.00,80.00){\line(1,0){70.00}}
\put(85.00,110.00){\line(1,-2){30.00}}
\put(85.74,28.01){\circle{2.14}}
\put(28.29,80.01){\circle*{3.40}}
\emline{97.90}{80.96}{1}{99.99}{80.03}{2}
\emline{97.90}{79.00}{3}{99.99}{80.01}{4}
\emline{114.92}{52.00}{5}{115.03}{50.04}{6}
\emline{113.47}{51.31}{7}{114.99}{49.97}{8}
\put(84.33,111.33){\circle*{3.40}}
\put(116.00,48.33){\circle*{3.40}}
\emline{86.67}{28.67}{9}{114.67}{46.67}{10}
\emline{85.67}{29.00}{11}{84.67}{109.33}{12}
\emline{85.00}{28.67}{13}{30.33}{78.00}{14}
\put(110.00,70.00){\makebox(0,0)[lb]{$\vec r$}}
\put(43.67,83.67){\makebox(0,0)[lb]{$\vec \rho$}}
\put(49.00,53.33){\makebox(0,0)[rt]{$\vec r_3$}}
\put(106.00,35.33){\makebox(0,0)[lb]{$\vec r_1$}}
\put(89.33,58.67){\makebox(0,0)[lc]{$\vec r_2$}}
\put(121.33,48.33){\makebox(0,0)[lc]{$N_1$}}
\put(84.00,116.33){\makebox(0,0)[cb]{$N_2$}}
\put(23.33,80.00){\makebox(0,0)[rc]{$N_3$}}
\put(75.60,63.99){\makebox(0,0)[lc]{$\vec p$}}
\emline{83.71}{105.90}{15}{84.66}{109.63}{16}
\emline{84.66}{109.63}{17}{85.66}{105.86}{18}
\emline{112.78}{46.30}{19}{114.93}{46.92}{20}
\emline{114.93}{46.92}{21}{113.59}{45.04}{22}
\emline{30.70}{76.04}{23}{29.53}{78.64}{24}
\emline{29.53}{78.64}{25}{32.43}{77.77}{26}
\put(85.33,23.67){\makebox(0,0)[ct]{$e$}}
\emline{85.00}{29.00}{27}{77.00}{80.00}{28}
\emline{76.30}{77.67}{29}{77.05}{80.00}{30}
\emline{77.05}{80.00}{31}{78.33}{77.97}{32}
\end{picture}
\end{figure}
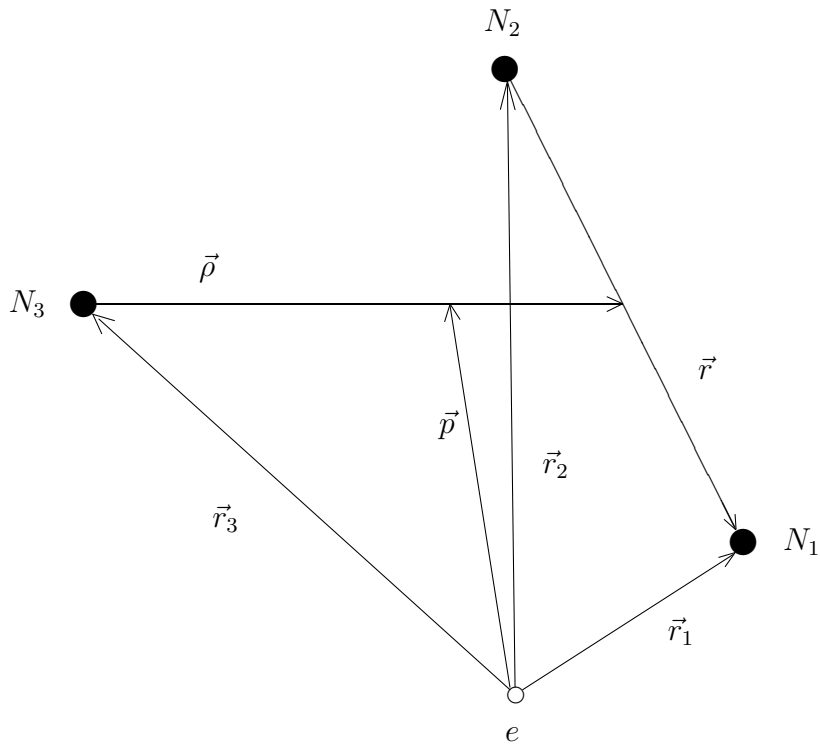

\newpage
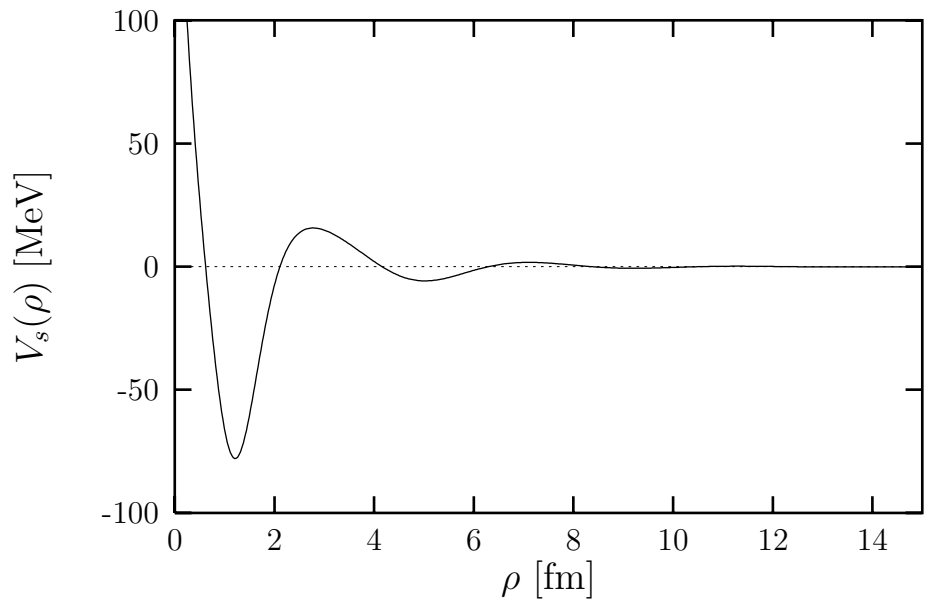
\begin{figure}
\caption{The $nd$ effective interaction obtained using the Marchenko
    inverse scattering  method.}
\vspace{2cm}
\setlength{\unitlength}{0.1bp}
\begin{picture}(3600,2160)(0,0)
\put(2008,1){\large\makebox(0,0){$\rho$ [fm]}}
\put(100,1180){%
\makebox(0,0)[b]{\large\shortstack{$V_s(\rho)$ [MeV]}}%
}
\put(3229,151){\makebox(0,0){14}}
\put(2854,151){\makebox(0,0){12}}
\put(2478,151){\makebox(0,0){10}}
\put(2102,151){\makebox(0,0){8}}
\put(1727,151){\makebox(0,0){6}}
\put(1351,151){\makebox(0,0){4}}
\put(976,151){\makebox(0,0){2}}
\put(600,151){\makebox(0,0){0}}
\put(540,2109){\makebox(0,0)[r]{100}}
\put(540,1644){\makebox(0,0)[r]{50}}
\put(540,1180){\makebox(0,0)[r]{0}}
\put(540,716){\makebox(0,0)[r]{-50}}
\put(540,251){\makebox(0,0)[r]{-100}}
\end{picture}

\end{figure}

\newpage
\begin{table}
\caption[]{Nonradiative,
   ${\cal R}_e=\langle{\cal R}(pde\rightarrow e{}^3{\rm He})\rangle$,
 and radiative,
 ${\cal R}_\gamma=\langle{\cal R}(pd\rightarrow{}^3{\rm He}\gamma)\rangle$,
capture rates
as functions of the plasma temperature $T_6$. The fourth column shows their
ratio for $n_e=100N_A \,{\rm cm}^{-3}$.
The temperature $T_6$ is in $10^6\,\,{}^{\circ}{\rm K}$ units, and the
reaction rates in ${\rm cm}^6{\rm mole}^{-2}{\rm sec}^{-1}$ and
${\rm cm}^3{\rm mole}^{-1}{\rm sec}^{-1}$ respectively. The data for
$\langle{\cal R}(pd\rightarrow{}^3{\rm He}\gamma)\rangle$ are taken from
Ref.\cite{fowler}.}
\begin{tabular}{|c|c|c|c||c|c|c|c|}
\hline
 $T_6$ & ${\cal R}_e$ & ${\cal R}_\gamma   $ & Ratio &
 $T_6$ & ${\cal R}_e$ & ${\cal R}_\gamma   $ &Ratio \\
\hline
 1  &  0.357 (-16)&  0.164 (-10) & 0.218 (-3) &  15  &  0.153 (-7)  &  0.132
(-1) & 0.116 (-3) \\
 2  &  0.423 (-13)&  0.228 (-7)  & 0.186 (-3) &  16  &  0.201 (-7)  &  0.176
(-1) & 0.114 (-3) \\
 3  &  0.124 (-11)&  0.741 (-6)  & 0.167 (-3) &  18  &  0.324 (-7)  &  0.292
(-1) & 0.111 (-3) \\
 4  &  0.103 (-10)&  0.658 (-5)  & 0.157 (-3) &  20  &  0.486 (-7)  &  0.453
(-1) & 0.107 (-3) \\
 5  &  0.459 (-10)&  0.309 (-4)  & 0.147 (-3) &  25  &  0.108 (-6)  &  0.109
     & 0.990 (-4) \\
 6  &  0.144 (-9) &  0.100 (-3)  & 0.144 (-3) &  30  &  0.198 (-6)  &  0.211
     & 0.938 (-4) \\
 7  &  0.354 (-9) &  0.255 (-3)  & 0.139 (-3) &  40  &  0.474 (-6)  &  0.557
     & 0.851 (-4) \\
 8  &  0.746 (-9) &  0.552 (-3)  & 0.135 (-3) &  50  &  0.863 (-6)  &  0.111
(+1) & 0.777 (-4) \\
 9  &  0.139 (-8) &  0.106 (-2)  & 0.131 (-3) &  60  &  0.135 (-5)  &  0.188
(+1) & 0.718 (-4) \\
 10 &  0.238 (-8) &  0.185 (-2)  & 0.129 (-3) &  70  &  0.191 (-5)  &  0.286
(+1) & 0.668 (-4) \\
 11 &  0.381 (-8) &  0.301 (-2)  & 0.127 (-3) &  80  &  0.253 (-5)  &  0.406
(+1) & 0.623 (-4) \\
 12 &  0.573 (-8) &  0.463 (-2)  & 0.124 (-3) &  90  &  0.318 (-5)  &  0.545
(+1) & 0.583 (-4) \\
 13 &  0.824 (-8) &  0.680 (-2)  & 0.121 (-3) &  100 &  0.387 (-5)  &  0.703
(+1) & 0.550 (-4) \\
 14 &  0.114 (-7) &  0.963 (-2)  & 0.118 (-3) &  200 &  0.106 (-4)  &  0.311
(+2) & 0.151 (-3) \\
\hline
\end{tabular}
\end{table}

\end{document}